\documentclass{article}
\usepackage{spconf,amsmath,amsfonts,amssymb,graphicx,hyperref,multirow,booktabs}
\usepackage{xcolor}
\usepackage{float} %
\usepackage{longtable}
\usepackage{soul} 
\sethlcolor{yellow!40}

\title{RPM-NET: RECIPROCAL POINT MLP NETWORK FOR UNKNOWN NETWORK SECURITY THREAT DETECTION}

\name{Jiachen Zhang$^\dagger$, Yueming Lu$^\dagger$\sthanks{Corresponding author~(Prof. Yueming Lu):~ymlu@bupt.edu.cn}, Fan Feng$^\ddagger$, Zhanfeng Wang$^\dagger$, Shengli Pan\textsuperscript{1}$^\dagger$, Daoqi Han$^\dagger$}

\address{%
$^\dagger$ School of Cyberspace Security, Beijing University of Posts and Telecommunications \\
$^\ddagger$ Guangxi Key Laboratory of Digital Infrastructure, \\Guangxi Zhuang Autonomous Region Information Center }

\begin{document}
\topmargin=0mm

%
\maketitle
\let\thefootnote\relax\footnotetext{Note: This version corrects a transcription error in Table 1 (ODIN's precision, recall, and f1 scores) compared to the conference proceedings.}
\begin{abstract}

       Effective detection of unknown network security threats in multi-class imbalanced environments is critical for maintaining cyberspace security. Current  methods focus on learning class representations but face challenges with unknown threat detection, class imbalance, and lack of interpretability, limiting their practical use. To address this, we propose RPM-Net, a novel framework that introduces reciprocal point mechanism to learn "non-class" representations for each known attack category, coupled with adversarial margin constraints that provide geometric interpretability for unknown threat detection. RPM-Net++ further enhances performance through Fisher discriminant regularization. Experimental results show that RPM-Net achieves superior performance across multiple metrics including F1-score, AUROC, and AUPR-OUT, significantly outperforming existing methods and offering practical value for real-world network security applications. Our code is available at: \url{https://github.com/chiachen-chang/RPM-Net}

\end{abstract}
\section{Introduction}
\label{sec:intro}
Network technology advancement and digital transformation have elevated network security to a critical challenge \cite{1,2}. Cyber attack techniques have become complex and diverse, expanding from conventional virus dissemination and denial of service attacks \cite{3} to sophisticated forms such as ransomware, supply chain attacks, and zero-day exploits \cite{4}. Attackers continuously refine their strategies, targeting victims with greater precision \cite{5}. 
\begin{figure}[htb]
\centerline{\includegraphics[width=8.5cm]{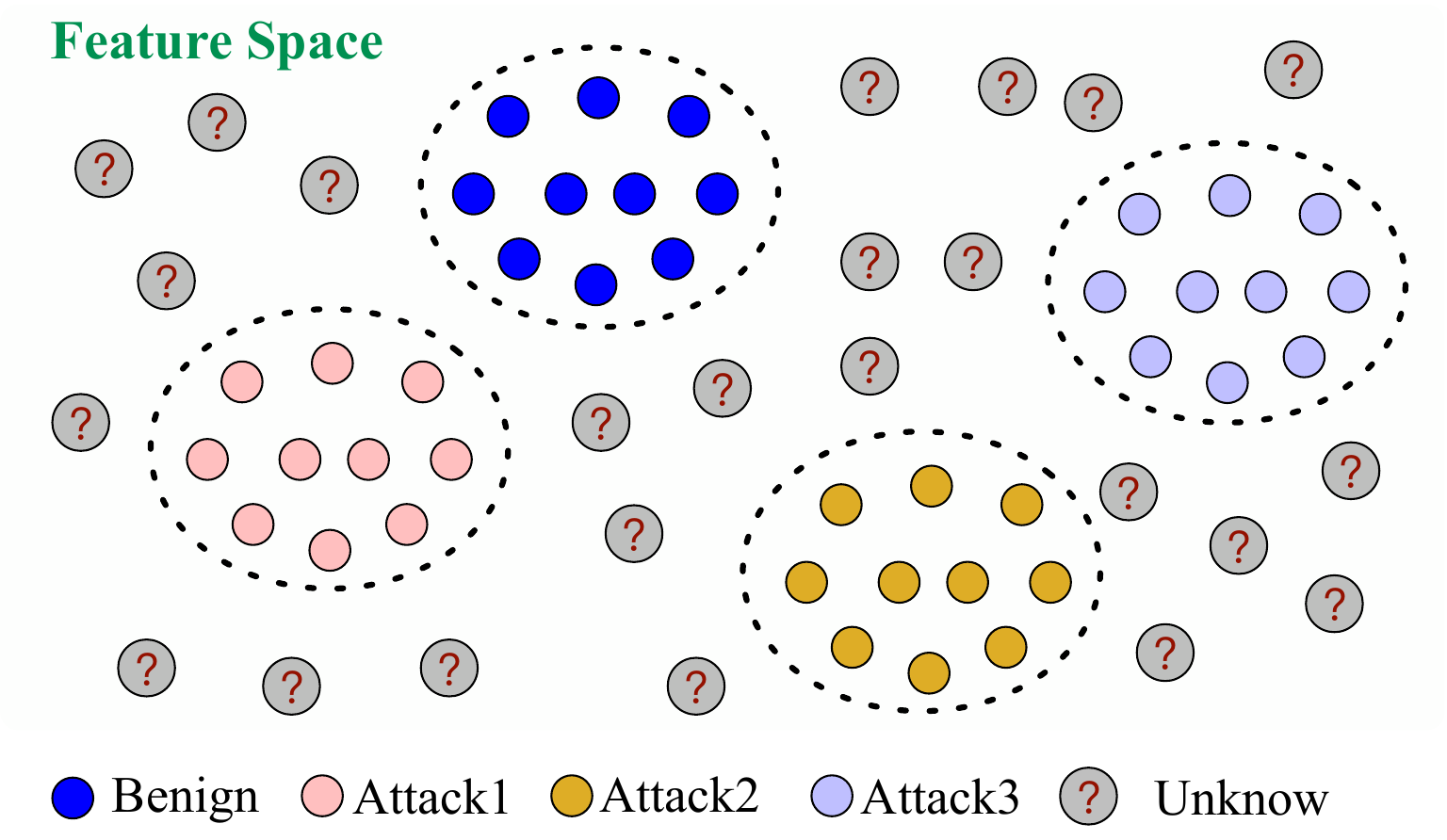}}
\caption{Illustration of open-set network threat detection.}
\label{fig:openset_concept}
\end{figure}
Conventional threat detection methods encounter unprecedented obstacles. The task involves accurate identification of known attack patterns and reliable detection of novel cyber threats \cite{han2024ecnet, farrukh2024ais}. This necessity has given rise to Open Set Recognition (OSR) \cite{yang2021conditional}. OSR effectively differentiates between known attacks and unfamiliar threats using only known attack samples during training. This capability offers early warning and threat identification essential for network security.

As illustrated in Figure~\ref{fig:openset_concept}, practical threat detection systems operate under the open set assumption, encountering both known attack patterns and previously unseen threats during deployment. This challenge requires mechanisms that maintain high classification accuracy on familiar attack types while reliably identifying novel threats beyond the training distribution. Scattered unknown patterns demonstrate that malicious activities can emerge anywhere, often where traditional closed-set classifiers make overconfident predictions.

Current open set identification techniques have notable theoretical and practical constraints. Firstly, most methods utilize simplistic binary classification, training only normal traffic or single attack types as positive samples \cite{6,7,8,9,10,11,dagmm}. This approach falls short in real network settings due to diverse attack types \cite{Shin2023OpenSR}, such as DDoS attacks, port scanning, brute force attacks, and malware propagation. Secondly, multi-class methods presume balanced class distribution, which conflicts with actual network scenarios where attack frequencies vary significantly. This class imbalance substantially impacts model generalization and detection accuracy \cite{Zhang2023MF2POSEMF}. Additionally, current methods lack efficient mechanisms to address relationships among known classes and construct feature spaces that naturally encompass unknown threat classes.

To address these challenges, we propose RPM-Net, a novel multi-category open set recognition framework that integrates Reciprocal Points to represent "non-class" spaces for each known attack category. By implementing Adversarial Margin Constraints, RPM-Net establishes adaptable boundary regions that push known attack types to the feature space periphery while creating central "open space" for unidentified threats, mitigating category imbalance without requiring unknown class samples during training. We further enhance RPM-Net with Fisher discriminant regularization (RPM-Net++) to improve intra-class compactness and inter-class separability. The main contributions include: (1) a reciprocal point mechanism enabling effective multi-class attack differentiation by learning non-certain class representations; (2) Fisher discriminant regularization enhancing intra-class compactness and inter-class separability; (3) adaptive boundary construction through margin constraints addressing feature space challenges in multi-class unbalanced environments; and (4) a unified training strategy utilizing only known attack data to establish open-set recognition feature space.

\begin{figure*}[htb]
\centerline{\includegraphics[width=17.6cm]{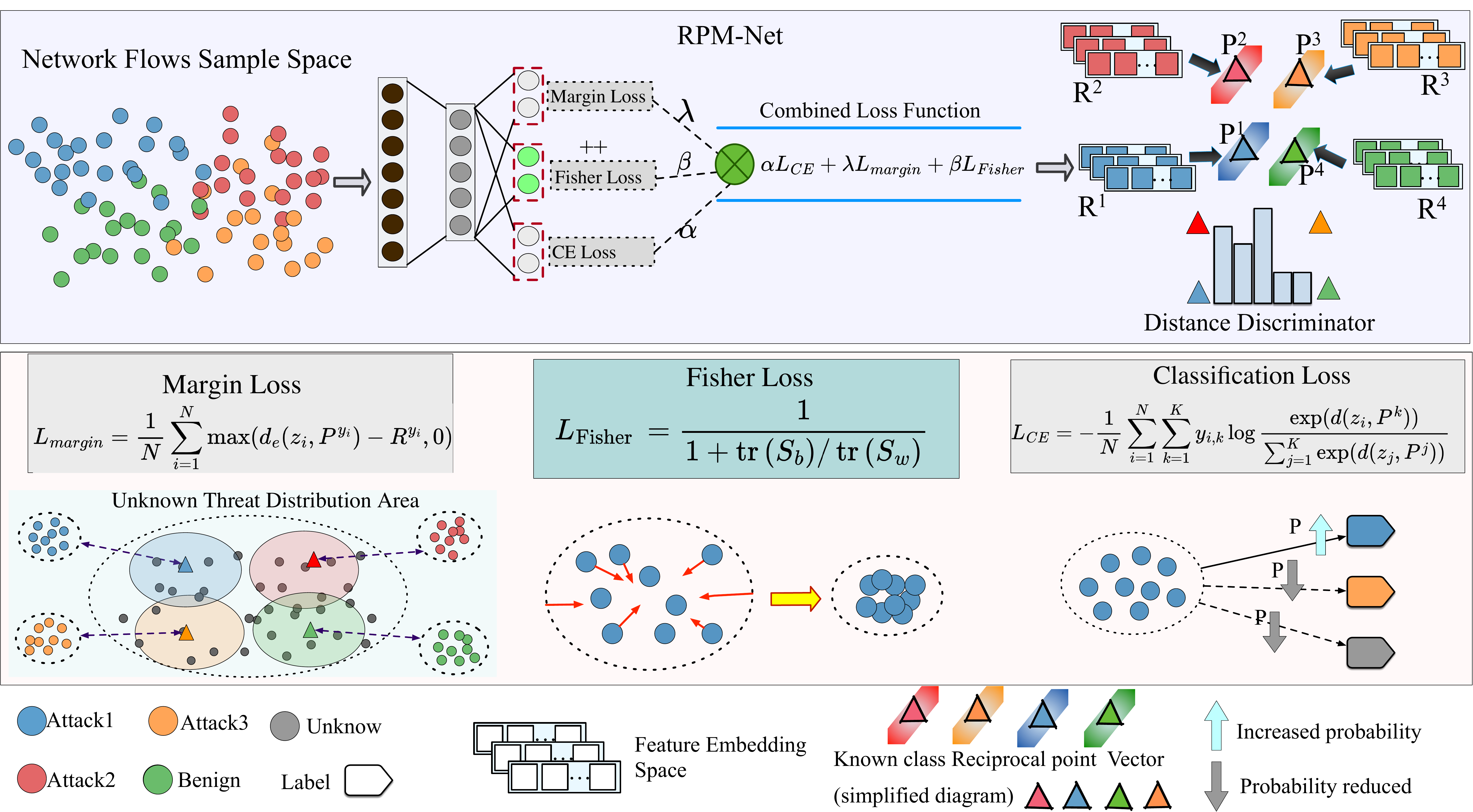}}
\caption{RPM-Net architecture and training process: (1)The top section illustrates the complete workflow from network flow samples through the neural network to reciprocal point space classification. (2)The bottom section details the three core loss functions: Margin Loss ensures separation between known classes and their reciprocal points, Fisher Loss enhances intra-class compactness and inter-class separability, and Classification Loss optimizes class probability distributions. The distance discriminator uses reciprocal point distances to classify new samples and detect unknown threats.}
\label{fig:rpmnet_architecture}
\end{figure*}
\section{Proposed method}
\label{proposed_method}
\textbf{2.1 Overall Architecture}:

The overall architecture of the proposed RPM-Net model is shown in Figure~\ref{fig:rpmnet_architecture}. RPM-Net  consists of four components: (1) feature extractor $\phi: \mathbb{R}^d \rightarrow \mathbb{R}^m$, (2) learnable reciprocal points $\{P^k\}_{k=1}^K$ for each known class, (3) adversarial margin constraints $\{R^k\}_{k=1}^K$, and (4) Fisher discriminant regularization (RPM-Net++ with it).
The feature extractor is implemented as a multi-layer perceptron with ReLU activations and dropout regularization:
\begin{align}
\phi(x) &= W_3 \cdot \text{ReLU}(\text{Dropout}(W_2 \cdot \text{ReLU}(\text{Dropout}(W_1 x \nonumber \\
&\quad+ b_1)) + b_2)) + b_3
\end{align}
where $W_i$ and $b_i$ are learnable weight matrices and bias vectors, and Dropout denotes dropout regularization.

\textbf{2.2 Reciprocal Point and Margin Constraints}:

Reciprocal points $P^k \in \mathbb{R}^m$ represent "what a class is not." For each known class $k$, the reciprocal point $P^k$ serves as the center of the feature space region that should not contain samples from class $k$.

The distance from an embedding $z = \phi(x)$ to reciprocal point $P^k$ is computed as:
\begin{equation}
d(z, P^k) = d_e(z, P^k) - d_c(z, P^k)
\end{equation}
where $d_e(z, P^k) = \|z - P^k\|_2^2 / m$ is the normalized Euclidean distance, and $d_c(z, P^k) = z^T P^k / (\|z\|_2 \|P^k\|_2)$ is the cosine similarity.

The classification logit for class $k$ is then:
\begin{equation}
\text{logit}_k(x) = \gamma \cdot d(z, P^k)
\end{equation}
where $\gamma$ is a scaling factor that controls the magnitude of the logits.

Learnable margin parameters $R^k > 0$ constrain known class samples to remain within distance $R^k$ from their corresponding reciprocal point:
\begin{equation}
L_{margin} = \frac{1}{N} \sum_{i=1}^N \max(d_e(z_i, P^{y_i}) - R^{y_i}, 0)
\end{equation}
where $d_e(z_i, P^{y_i})$ is the normalized Euclidean distance from sample $i$ to its corresponding reciprocal point.

This constraint prevents feature space explosion and creates boundaries that accommodate unknown classes in the central region.

\textbf{2.3 Fisher Regularization and Training Objective}:

Fisher discriminant regularization maximizes the ratio of inter-class scatter to intra-class scatter. For embeddings $\{z_i\}$ with corresponding labels $\{y_i\}$, we compute the within-class scatter and between-class scatter:

\begin{equation}
S_w = \sum_{k=1}^K \sum_{i: y_i=k} \|z_i - \mu_k\|_2^2
\end{equation}

\begin{equation}
S_b = \sum_{k=1}^K n_k \|\mu_k - \mu\|_2^2
\end{equation}

where $\mu_k$ is the mean embedding of class $k$, $\mu$ is the global mean, and $n_k$ is the number of samples in class $k$. The Fisher discriminant criterion maximizes $S_b/S_w$, reformulated as a loss function:

\begin{equation}
L_{Fisher} = \frac{1}{1 + S_b/S_w}
\end{equation}

The overall training loss combines three objectives:
\begin{equation}
L_{total} = \alpha L_{CE} + \lambda L_{margin} + \beta L_{Fisher}
\end{equation}
where $L_{CE}$ is cross-entropy loss using reciprocal point distances, $L_{margin}$ enforces margin constraints, and $L_{Fisher}$ promotes intra-class compactness and inter-class separability. Hyperparameters $\alpha = 1.0$, $\lambda = 1.0$, and $\beta = 1.0$.

During training, reciprocal points and margins adapt to the data distribution, with known classes pushed toward the periphery and unknown regions forming in the center.

\textbf{2.4 Inference and Unknown Detection}:

We compute the maximum reciprocal point distance:
\begin{equation}
s(x) = \max_{k=1}^K d(\phi(x), P^k)
\end{equation}

For classification: $\hat{y} = \arg\max_{k=1}^K d(\phi(x), P^k)$

For unknown detection: if $s(x) < \tau$, the sample is classified as unknown. Threshold $\tau$ is determined using validation data.

\begin{table*}[h]
\centering
\caption{Performance comparison on CICIDS2017 and UNSW-NB15 datasets.
Precision, Recall, and F1-score are macro-averaged, and the best performance is marked in bold.}
\label{tab:results}
\renewcommand{\arraystretch}{0.93}
\small

\begin{tabular}{@{}l@{\hspace{0.96cm}}l@{\hspace{0.96cm}}c@{\hspace{0.96cm}}c@{\hspace{0.96cm}}c@{\hspace{0.96cm}}c@{\hspace{0.96cm}}c@{\hspace{0.96cm}}c@{}}
\toprule
\textbf{Dataset} & \textbf{Method} & \textbf{Precision} & \textbf{Recall} & \textbf{F1-Score} & \textbf{AUROC} & \textbf{AUPR-IN} & \textbf{AUPR-OUT} \\
\midrule
\multirow{5}{*}{CICIDS2017} 
& Baseline & \textbf{0.9996} & \textbf{0.9995} & \textbf{0.9996} & 0.7069 & 0.9732 & 0.1046 \\
& EVM      & 0.9806 & 0.5538 & 0.6082 & 0.9600 & 0.9986 & 0.2974 \\
& OCN      & 0.9946 & 0.9967 & 0.9956 & 0.9057 & 0.9968 & 0.2884 \\
& ODIN     & 0.9966 & 0.9582 & 0.9765 & 0.7354 & 0.9726 & 0.1651 \\
& RPM-Net   & 0.9987 & 0.9987 & 0.9987 & \textbf{0.9601} & \textbf{0.9989} & \textbf{0.6523} \\
\midrule
\multirow{5}{*}{UNSW-NB15}
& Baseline & 0.6626 & 0.5912 & 0.5473 & 0.7867 & 0.7928 & 0.7867 \\
& EVM      & 0.8485 & 0.5155 & 0.6259 & 0.8300 & 0.8370 & 0.8322 \\
& OCN      & 0.7826 & 0.7773 & 0.7605 & 0.7815 & 0.7515 & 0.7814 \\
& ODIN     & \textbf{0.8533} & 0.6598 & 0.7302 & 0.8172 & 0.7748 & 0.8277 \\
& RPM-Net   & 0.8022 & \textbf{0.8053} & \textbf{0.7950} & \textbf{0.8675} & \textbf{0.8511} & \textbf{0.8555} \\
\bottomrule
\end{tabular}
\end{table*}


\begin{table*}[h]
\centering
\caption{Ablation study: Impact of Fisher discriminant regularization, the best performance is marked in bold.}
\label{tab:ablation}
\renewcommand{\arraystretch}{0.93}  
\small
\begin{tabular}{@{}l@{\hspace{0.92cm}}l@{\hspace{0.92cm}}c@{\hspace{0.92cm}}c@{\hspace{0.92cm}}c@{\hspace{0.92cm}}c@{\hspace{0.92cm}}c@{\hspace{0.92cm}}c@{}}
\toprule
\textbf{Dataset} & \textbf{Method} & \textbf{Precision} & \textbf{Recall} & \textbf{F1-Score} & \textbf{AUROC} & \textbf{AUPR-IN} & \textbf{AUPR-OUT} \\
\midrule
\multirow{2}{*}{CICIDS2017} 
& RPM-Net & $\mathbf{0.9987}$ & $\mathbf{0.9987}$ & $\mathbf{0.9987}$ & 0.9601 & 0.9989 & 0.6523 \\
& RPM\text{-}Net++ & 0.9981 & 0.9977 & 0.9979 & $\mathbf{0.9735}$ & 0.9989 & $\mathbf{0.6711}$ \\
\midrule
\multirow{2}{*}{UNSW-NB15}
& RPM-Net & 0.8022 & 0.8053 & 0.7950 & 0.8675 & 0.8511 & 0.8555 \\
& RPM\text{-}Net++ & $\mathbf{0.8043}$ & $\mathbf{0.8072}$ & $\mathbf{0.7955}$ & $\mathbf{0.8850}$ & $\mathbf{0.8913}$ & $\mathbf{0.8664}$ \\
\bottomrule
\end{tabular}
\end{table*}
\section{Experiments and analysis}
\label{experiment_analysis}
We evaluate RPM-Net on CICIDS2017 \cite{cic2017} and UNSW-NB15 \cite{unsw} datasets. CICIDS2017 contains 5 known classes (Benign, DDoS, DoS Hulk, PortScan, FTP-Patator), 2 validation classes, and 4 unknown test classes. UNSW-NB15 comprises 6 known classes (Benign, Analysis, Backdoor, DoS, Generic, Worms), 1 validation class, and 3 unknown classes. Data is preprocessed with z-score normalization and split 8:2 for training/testing. We compare against Baseline \cite{hendrycks2017baseline}, ODIN \cite{odin}, OCN \cite{ocn}, and EVM \cite{geng2020recent} methods.

\textbf{3.1 Results and Analysis}:

Table~\ref{tab:results} shows RPM-Net's performance across both datasets. On CICIDS2017, RPM-Net achieves a macro F1-score of 0.9987 for known-class classification and an AUPR-OUT of 0.6523 for unknown detection, significantly outperforming EVM (0.2974). The high AUPR-OUT indicates effective discrimination between known and unknown network traffic. On UNSW-NB15, RPM-Net maintains strong performance with an F1-score of 0.7950 and an AUPR-OUT of 0.8555, outperforming other methods. The consistent results across datasets demonstrate the generalizability of RPM-Net.

\textbf{3.2 Ablation Study}: Table \ref{tab:ablation} compares RPM-Net (base method) and RPM-Net++ (with Fisher regularization). On CICIDS2017, Fisher regularization improves AUROC from 0.9601 to 0.9735 (+1.40\%) and AUPR-OUT from 0.6523 to 0.6711 (+2.88\%). On UNSW-NB15, it enhances AUROC from 0.8675 to 0.8850 (+2.02\%), AUPR-IN from 0.8511 to 0.8913 (+4.72\%), and AUPR-OUT from 0.8555 to 0.8664 (+1.27\%). These improvements demonstrate that Fisher discriminant regularization effectively enhances intra-class compactness and inter-class separability, leading to better discrimination between known and unknown classes. The consistent gains across both datasets validate the synergistic effect of combining reciprocal points, margin constraints, and Fisher regularization.
\begin{figure}[htb]
\centerline{\includegraphics[width=9cm]{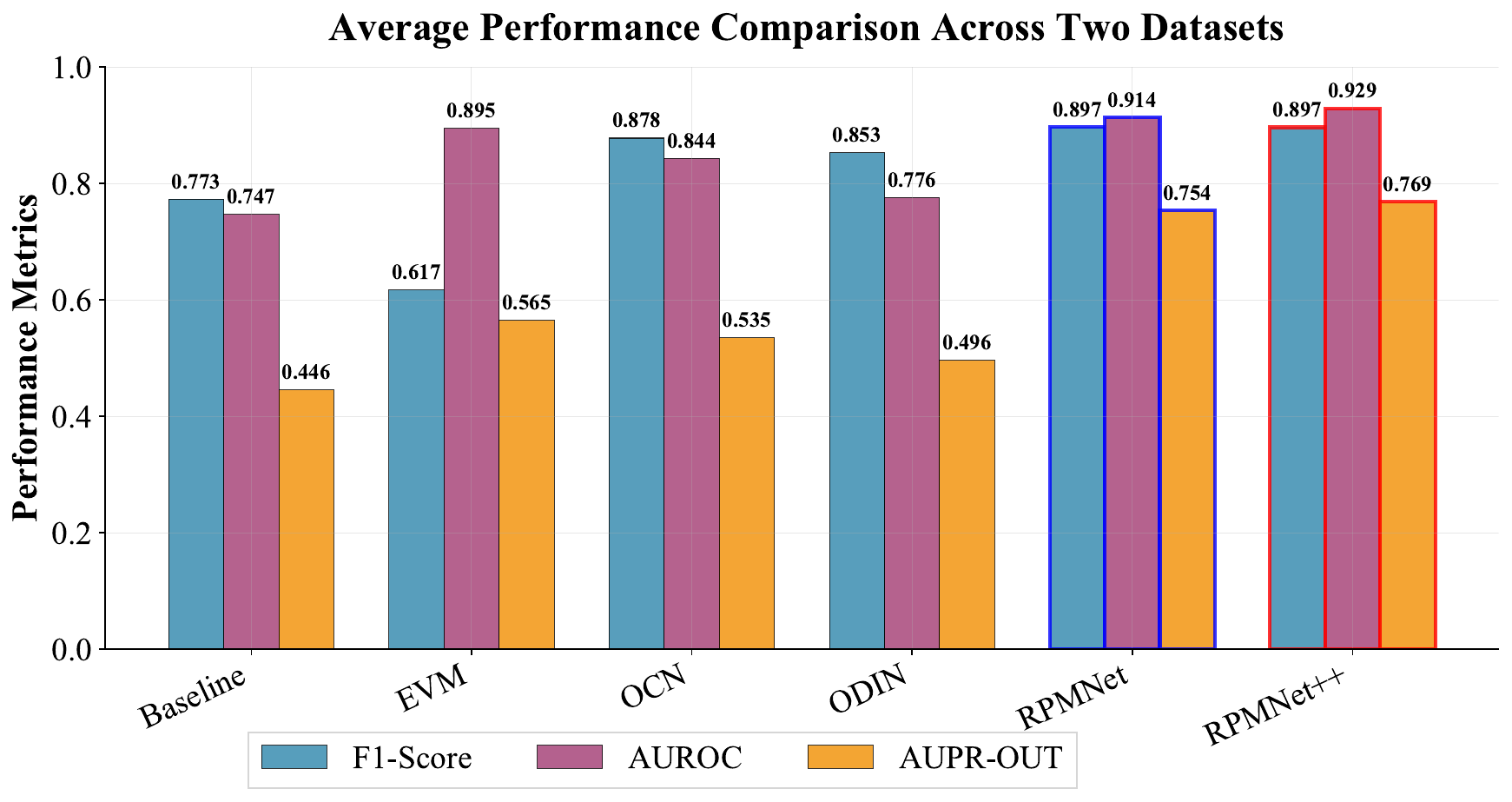}}
\caption{Average performance comparison across two datasets showing F1-Score, AUROC, and AUPR-OUT metrics for different methods.}
\label{fig:performance_comparison}
\end{figure}

\textbf{3.3 Comprehensive Performance Analysis}: Figure~\ref{fig:performance_comparison} shows average performance across both datasets. RPM-Net achieves the highest performance across all metrics, with particularly pronounced improvements in AUPR-OUT, highlighting superior unknown threat detection capability. The results demonstrate that RPM-Net effectively addresses open set recognition challenges in network security, maintaining high accuracy for known attacks while identifying novel threats in real-world scenarios.

\section{Conclusion}
\label{conclusion}
In this paper, we propose RPM-Net for network security threat detection, which includes reciprocal point mechanism, adversarial margin constraints, and fisher discriminant regularization(RPM-Net++). The reciprocal point mechanism learns "non-class" representations for each known attack category, while margin constraints create bounded feature spaces naturally accommodating unknown classes. Experiments show RPM-Net++ achieves superior performance with 99.79\% F1-score and 67.11\% AUPR-OUT on CICIDS2017, and 79.55\% F1-score and 86.64\% AUPR-OUT on UNSW-NB15, significantly outperforming baseline methods. The framework's ability to handle class imbalance without requiring unknown class samples during training makes it suitable for real-world network security applications. Future work will explore extensions to streaming data scenarios and applications to other security domains.

\section{Acknowledgment}
This work was supported by the Science and Technology Projects of Xizang Autonomous Region, China (Grant No. XZ202501ZY0026) and the Open Project Program of Guangxi Key Laboratory of Digital Infrastructure (Grant No. GXDIOP2024018).

\bibliographystyle{IEEEbib}
\bibliography{refs}

@article{1,
  title={Operating system network security enhancement scheme based on trusted storage},
  author={Qi, Longyun and Lv, Xiaoliang and Sun, Lianwen and Yao, Tianle and Yu, Jianye and Wang, Lei},
  journal={Intelligent and Converged Networks},
  volume={4},
  number={2},
  pages={127--141},
  year={2023},
  publisher={TUP}
}

@article{2,
  title={Multi-step attack detection based on pre-trained hidden Markov models},
  author={Zhang, Xu and Wu, Ting and Zheng, Qiuhua and Zhai, Liang and Hu, Haizhong and Yin, Weihao and Zeng, Yingpei and Cheng, Chuanhui},
  journal={Sensors},
  volume={22},
  number={8},
  pages={2874},
  year={2022},
  publisher={MDPI}
}

@article{3,
  title={Security Engineering of Patient-Centered Health Care Information Systems in Peer-to-Peer Environments: Systematic Review},
  author={Imrana Abdullahi Yari and Tobias Dehling and Felix Kluge and Juergen Geck and Ali Sunyaev and Bjoern M. Eskofier},
  journal={Journal of Medical Internet Research},
  year={2020},
  volume={23},
  url={https://api.semanticscholar.org/CorpusID:244115481}
}

@article{4,
  title={Advanced Persistent Threat (APT) and intrusion detection evaluation dataset for linux systems 2024},
  author={Karim, Syed Sohaib and Afzal, Mehreen and Iqbal, Waseem and Al Abri, Dawood},
  journal={Data in Brief},
  volume={54},
  pages={110290},
  year={2024},
  publisher={Elsevier}
}

@article{5,
  title={A sequential deep learning framework for a robust and resilient network intrusion detection system},
  author={Hore, Soumyadeep and Ghadermazi, Jalal and Shah, Ankit and Bastian, Nathaniel D},
  journal={Computers \& Security},
  volume={144},
  pages={103928},
  year={2024},
  publisher={Elsevier}
}

@article{han2024ecnet,
  title={ECNet: Robust malicious network traffic detection with multi-view feature and confidence mechanism},
  author={Han, Xueying and Liu, Song and Liu, Junrong and Jiang, Bo and Lu, Zhigang and Liu, Baoxu},
  journal={IEEE Transactions on Information Forensics and Security},
  year={2024},
  publisher={IEEE}
}

@article{farrukh2024ais,
  title={Ais-nids: An intelligent and self-sustaining network intrusion detection system},
  author={Farrukh, Yasir Ali and Wali, Syed and Khan, Irfan and Bastian, Nathaniel D},
  journal={Computers \& Security},
  volume={144},
  pages={103982},
  year={2024},
  publisher={Elsevier}
}

@article{yang2021conditional,
  title={Conditional variational auto-encoder and extreme value theory aided two-stage learning approach for intelligent fine-grained known/unknown intrusion detection},
  author={Yang, Jian and Chen, Xiang and Chen, Shuangwu and Jiang, Xiaofeng and Tan, Xiaobin},
  journal={IEEE Transactions on Information Forensics and Security},
  volume={16},
  pages={3538--3553},
  year={2021},
  publisher={IEEE}
}

@article{Zhang2023MF2POSEMF,
  title={MF2POSE: Multi-task Feature Fusion Pseudo-Siamese Network for intrusion detection using Category-distance Promotion Loss},
  author={Jiawei Zhang and Rui Chen and Yanchun Zhang and Weihong Han and Zhaoquan Gu and Shuqiang Yang and Yongquan Fu},
  journal={Knowl. Based Syst.},
  year={2023},
  volume={283},
  pages={111110},
  url={https://api.semanticscholar.org/CorpusID:264510626}
}

@inproceedings{dagmm,
  title={Deep Autoencoding Gaussian Mixture Model for Unsupervised Anomaly Detection},
  author={Bo Zong and Qi Song and Martin Renqiang Min and Wei Cheng and Cristian Lumezanu and Dae-ki Cho and Haifeng Chen},
  booktitle={International Conference on Learning Representations},
  year={2018},
  url={https://api.semanticscholar.org/CorpusID:51805340}
}

@article{Shin2023OpenSR,
  title={Open Set Recognition With Dissimilarity Weight for Unknown Attack Detection},
  author={Gun-Yoon Shin and Dong-Wook Kim and Myung-Mook Han},
  journal={IEEE Access},
  year={2023},
  volume={11},
  pages={102381-102390},
  url={https://api.semanticscholar.org/CorpusID:258811401}
}

@article{6,
  title={Enhancing IoT Network Security: Unveiling the Power of Self-Supervised Learning against DDoS Attacks},
  author={Josue Genaro Almaraz-Rivera and Jos{\'e} Antonio Cantoral-Ceballos and Juan Felipe Botero},
  journal={Sensors (Basel, Switzerland)},
  year={2023},
  volume={23},
  url={https://api.semanticscholar.org/CorpusID:264503788}
}

@article{7,
  title={Few-Shot network intrusion detection based on prototypical capsule network with attention mechanism},
  author={Handi Sun and Liang Wan and Mengying Liu and Bo Wang},
  journal={PLOS ONE},
  year={2023},
  volume={18},
  url={https://api.semanticscholar.org/CorpusID:258237238}
}

@article{8,
  title={Distributed Denial of Service Attack Detection in Network Traffic Using Deep Learning Algorithm},
  author={Mahrukh Ramzan and Muhammad Shoaib and Ayesha Altaf and Shazia Arshad and Faiza Iqbal and {\'A}ngel Kuc Castilla and Imran Ashraf},
  journal={Sensors (Basel, Switzerland)},
  year={2023},
  volume={23},
  url={https://api.semanticscholar.org/CorpusID:264470054}
}

@article{9,
  title={A Convolutional Neural Network for Improved Anomaly-Based Network Intrusion Detection},
  author={Isra M. Al-Turaiki and Najwa Altwaijry},
  journal={Big Data},
  year={2021},
  volume={9},
  pages={233 - 252},
  url={https://api.semanticscholar.org/CorpusID:235470654}
}

@article{10,
  title={Network traffic classification for data fusion: A survey},
  author={Jingjing Zhao and Xuyang Jing and Zheng Yan and Witold Pedrycz},
  journal={Inf. Fusion},
  year={2021},
  volume={72},
  pages={22-47},
  url={https://api.semanticscholar.org/CorpusID:233580368}
}

@article{11,
  title={SMOTE-DRNN: A Deep Learning Algorithm for Botnet Detection in the Internet-of-Things Networks},
  author={Segun I. Popoola and Bamidele Adebisi and Ruth Ande and Mohammad Hammoudeh and Kelvin O. O. Anoh and Atayero},
  journal={Sensors (Basel, Switzerland)},
  year={2021},
  volume={21},
  url={https://api.semanticscholar.org/CorpusID:233462268}
}

@article{unsw,
  title={Poisoning and Evasion: Deep Learning-Based NIDS under Adversarial Attacks},
  author={Hesamodin Mohammadian and Arash Habibi Lashkari and Ali A. Ghorbani},
  journal={2024 21st Annual International Conference on Privacy, Security and Trust (PST)},
  year={2024},
  pages={1-9},
  url={https://api.semanticscholar.org/CorpusID:274815927}
}

@inproceedings{cic2017,
  title={Toward Generating a New Intrusion Detection Dataset and Intrusion Traffic Characterization},
  author={Iman Sharafaldin and Arash Habibi Lashkari and Ali A. Ghorbani},
  booktitle={International Conference on Information Systems Security and Privacy},
  year={2018},
  url={https://api.semanticscholar.org/CorpusID:4707749}
}

@inproceedings{odin,
  title={Enhancing The Reliability of Out-of-distribution Image Detection in Neural Networks},
  author={Liang, Shiyu and Li, Yixuan and Srikant, R},
  booktitle={International Conference on Learning Representations},
  year={2018}
}

@inproceedings{hendrycks2017baseline,
  title={A Baseline for Detecting Misclassified and Out-of-Distribution Examples in Neural Networks},
  author={Hendrycks, Dan and Gimpel, Kevin},
  booktitle={International Conference on Learning Representations},
  year={2017}
}

@article{geng2020recent,
  title={Recent advances in open set recognition: A survey},
  author={Geng, Chuanxing and Huang, Sheng-jun and Chen, Songcan},
  journal={IEEE transactions on pattern analysis and machine intelligence},
  volume={43},
  number={10},
  pages={3614--3631},
  year={2020},
  publisher={IEEE}
}

@article{ocn,
  title={A scalable network intrusion detection system towards detecting, discovering, and learning unknown attacks},
  author={Zhao Zhang and Yong Zhang and Da Guo and Mei Song},
  journal={International Journal of Machine Learning and Cybernetics},
  year={2021},
  volume={12},
  pages={1649 - 1665},
  url={https://api.semanticscholar.org/CorpusID:234280366}
}

@article{auroc,
  title={Out-of-distribution detection by cross-class vicinity distribution of in-distribution data},
  author={Zhao, Zhilin and Cao, Longbing and Lin, Kun-Yu},
  journal={IEEE Transactions on Neural Networks and Learning Systems},
  volume={35},
  number={10},
  pages={13777--13788},
  year={2023},
  publisher={IEEE}
}

\end{document}